\documentclass[pra,aps,twocolumn,showpacs]{revtex4}

\usepackage{amsmath}
\usepackage{amssymb}
\usepackage{graphicx}

\newcommand{\del} {\delta(\tau)}

\newcommand{\X} {\Upsilon}

\newcommand{\F} {{\cal \hat F}}
\renewcommand{\H} {{\cal \hat H}}

\begin{document}
\author{Thomas Salzburger$^1$, Peter Domokos$^2$, and Helmut Ritsch$^1$}
\title{Theory of a single-atom laser including light forces} 
\affiliation{$^1$Institute for Theoretical Physics, University of Innsbruck, A-6020 Innsbruck, Technikerstraße 25, Austria\\
$^2$Research Institute for Solid State Physics and Optics, P.\ O.\ Box 49, H-1525 Budapest, Hungary}

\begin{abstract}
We study a single incoherently pumped atom moving within an optical high-$Q$ resonator in the strong coupling regime.
Using a semiclassical description for the atom and field dynamics, we derive a closed system of differential equations to describe this coupled atom-field dynamics.
For sufficiently strong pumping the system starts lasing when the atom gets close to a field antinode, and the associated light forces provide for self-trapping of the atom.
For a cavity mode blue detuned with respect to the atomic transition frequency this is combined with cavity induced motional cooling allowing for long term steady-state operation of such a laser.
The analytical results for temperature and field statistics agree well with our earlier predictions based on Quantum Monte Carlo simulations.
We find sub-Doppler temperatures that decrease with gain and coupling strength and can even go beyond the limit of passive cavity cooling.
Besides demonstrating the importance of light forces in single-atom lasers, this result also gives strong evidence to enhance laser cooling through stimulated emission in resonators.
\end{abstract}

\pacs{32.80.Pj, 42.50.Vk, 42.50.Lc}

\maketitle

\section{Introduction}
A single incoherently pumped atom within a high-$Q$ optical resonator constitutes the smallest and conceptually simplest conceivable laser \cite{Salzburger04}.
On the one hand, this model is particularly interesting as it allows to theoretically study important aspects of laser physics analytically \cite{Briegel,Pellizzari}, while, on the other hand, there is great technological interest in very small tailored coherent light sources.
Already two decades ago laser like systems have been set up in the microwave regime \cite{Walther,Haroche}.
With the tremendous recent progress in laser cooling and micro cavity technology such systems have now indeed been experimentally realized \cite{Feld,Lange,Becher,Kimble} in the optical regime.
This requires ultracold atoms trapped in a rather small volume between mirrors of extremely high quality.
In this regime the light forces induced by the cavity field on the atom get important and have to be accounted for.
In a first approximation these forces are detrimental by heating the atomic motion and limiting the operation time of the system.
This heating is less problematic for a trapped ion, but here the spatial requirements for the ion trap prevent to reach the strong coupling regime.
In optical cavity QED setups such heating is significant and strongly shortens the interaction time \cite{KimblePRA}.

In a recent paper we have shown that for carefully chosen operating conditions one can reverse the detrimental effect of heating and combine gain with optical cooling and trapping \cite{Salzburger04}.
In this way the laser field generated by the atom can be used to simultaneously trap and cool the atom, which leads to a self-sustained sustained laser operation.
Interestingly, the temperature attained by the atom can be even lower than for free space Doppler cooling at comparable parameter values.
This stems from the strong nonlinear dependence of the field intensity on the atomic position, which gives new prospects of developing a novel laser cooling method enhanced by stimulated emission.
The numerical results obtained in Ref.\ \cite{Salzburger04} clearly demonstrate the significance and potential usefulness of light forces for single-atom lasing \cite{KimblePRA}, but they only provide little insight in the basic physics going on to facilitate this interesting behavior.     

In this work we develop a systematic semiclassical description of such a single-atom laser in the spirit of the successful models developed for cavity cooling \cite{Domokos01}.
Approximating the atomic center of mass motion by a point particle, we are able to find a set of coupled equations for the internal atomic dynamics and field evolution parameterized by the atomic position.
In contrast to laser cooling we cannot adiabatically eliminate the upper atomic state as we need an inverted atom for gain.
Nevertheless, by use of a factorization approximation for higher order atom-field expectation values, we still can derive a closed system of coupled equations for the combined dynamics.
As a proven but simple model for pumping we use an inverted heat bath approach developed by Haken several decades ago \cite{Haken}.
From these internal dynamical equations we are then able to derive the average light potential as well as friction and momentum diffusion coefficients for the atomic center of mass motion in an analytic form.
This allows detailed studies of atomic motion and laser field evolution that can be checked and compared to quantum Monte Carlo simulations for selected parameters.                  

\section{Model}
Let us consider a single inverted two-level atom freely moving in the field of an optical resonator with high finesse.
The relevant mode with frequency $\omega_c = kc$ is detuned from the atomic transition frequency $\omega_a$ by $\Delta = \omega_c - \omega_a$.
The dipole--resonator-field interaction Hamiltonian in the rotating wave and the electric dipole approximation ($\hbar = 1$) then reads
\begin{equation}
  \H = - \Delta\sigma_+\sigma_- + i G\left(a^\dagger \sigma_- - \sigma_+a \right)
\end{equation}
which is written in a frame rotating at the cavity resonance frequency.
Here $\sigma_-$ ($\sigma_+$) and $a$ denote, respectively, the atomic lowering (raising) operator and the bosonic field operator for the cavity mode while the position dependent atom-field coupling is given by $G = g\cos(kx)$.


Both the field and the atom are coupled to the environment, which is modeled by Markovian decay processes with rates $2\kappa$ (photon loss via the mirrors) and $2\gamma$ (spontaneous emission).
Using standard techniques of quantum optics, we can derive the following master equation:

\begin{equation}
  \frac{d}{dt}\hat{\varrho} = -i\bigl[{\H},\hat{\varrho}\bigr] + {\cal L}_\kappa\hat{\varrho}
  + {\cal L}_\gamma\hat{\varrho}\,.
  \label{eq:rho}
\end{equation}
It describes the time evolution of the resulting open system including decay of the resonator mode
$({\cal L}_\kappa\hat{\varrho})$ and the atomic upper state $({\cal L}_\gamma\hat{\varrho})$.

In order to feed energy into the system, the atom is driven externally by incoherent excitation at rate $2\nu$.
As a simple but still quantum mechanically consistent way to incorporate such a pumping mechanism, we model it by inverse spontaneous emission.
This proven method has been introduced already in the early quantum models of lasing \cite{Haken}, and we simply have to add a corresponding Liouvillian term to Eq.\ \eqref{eq:rho}:
\begin{equation}
  {\cal L}_\nu\hat{\varrho} = \nu(2\sigma_+\hat{\varrho}\sigma_- -\sigma_-\sigma_+\hat{\varrho} - \hat{\varrho}\sigma_-\sigma_+)\,.
\end{equation}
In fact it is largely equivalent to pumping from the ground state to a third intermediate level, with a fast incoherent decay to the upper atomic state.

The above master equation is equivalent to the following set of Heisenberg-Langevin equations \cite{Gardiner_book}:
\begin{subequations}
  \label{eq:hle_0}
  \begin{gather}
    \dot{a} = - \kappa a + G\sigma_- + \xi_\alpha \\
    \dot{\sigma}_- = (i\Delta - \gamma - \nu)\sigma_- + G\sigma_za + \xi_\sigma \\
    \dot{\sigma}_z = -2(\gamma+\nu)\sigma_z - 2G(a^\dagger\sigma_- + \sigma_+a) + 2(\nu-\gamma) + \xi_z\,.
  \end{gather}
\end{subequations}
Here we have introduced noise operators originating from the coupling of the system to the environment.
While their expecation values vanish when the environment is a $T=0$ heat bath, their nonvanishing correlation functions are given by:
\begin{subequations}
  \begin{gather}
    \langle \xi_\alpha(t)\xi_\alpha^\dagger(t-\tau)\rangle = 2\kappa\delta(\tau)\\
    \langle \xi_\sigma(t)\xi_\sigma^\dagger(t-\tau)\rangle = 2\gamma\delta(\tau)\\
    \langle \xi_\sigma^\dagger(t)\xi_\sigma(t-\tau)\rangle = 2\nu\delta(\tau)\\
    \langle \xi_z(t)\xi_z(t-\tau)\rangle = \bigl(4(\gamma-\nu)\langle\sigma_z\rangle + 4(\gamma+\nu)\bigr)\delta(\tau)\,.
  \end{gather}
\end{subequations}

In this paper we assume that the kinetic atomic temperature stays well above the recoil limit $k_B T_{\mbox{\tiny rec}} = \hbar^2 k^2/(2m)$, which of course has to be self-consistently checked at the end.
This allows for a so called semiclassical approximation, where the particle's position and momentum are treated classically and enter the equations for the field and internal atomic dynamics simply as real parameters $x$ and $p$.
A systematic way to derive these equations for an atom in a cavity field is, e.g., presented in Ref.\ \cite{Domokos01}. 
In this approximation simulations of atomic trajectories will then be governed by the following Langevin-type equations
\begin{subequations}
  \label{eq:ext}
  \begin{gather}
    \dot{x} = p/m\\
    \dot{p} = F + \xi\, ,
\end{gather}
\end{subequations}
where $F$ denotes the average force acting on the atom and $\xi$ is a noise term giving rise to momentum diffusion.
Both of these values have to be calculated from the corresponding solution of the master equation Eq.\ \eqref{eq:rho}. 

\section{lamb semiclassical model}

Let us first try to get some qualitative insight into the dynamics of our system and neglect all the noise terms $\xi_i$ and replace all operators by c-numbers, i.e: $\langle a\rangle = \alpha$, $\langle \sigma_- \rangle = s$, and $\langle\sigma_z\rangle = z$.
Hence, we get the following set of coupled differential equations:
\begin{subequations}
  \label{eq:lamb}
  \begin{gather}
    \dot{\alpha} = -\kappa\alpha + Gs \label{eq:lamb_alpha}\\
    \dot{s} = (i\Delta-\gamma-\nu)s + Gz\alpha \\
    \label{eq:lamb_z} \dot{z} = -2(\gamma+\nu)z - 2G(\alpha^*s+s^*\alpha) + 2(\nu-\gamma) \, .
  \end{gather}
\end{subequations}
This can be readily solved for the steady state for an atom at fixed position.
Although any operator and noise correlations are neglected in this ``Lamb-type'' model, we still get some vital insights into the dynamical characteristics of our system.
As an immediate consequence, Eqs.\ \eqref{eq:lamb} imply the  continuity equation
\begin{equation}
  \label{eq:cont} \nu (1-P) = \gamma P + \kappa N
\end{equation}
which describes the energy balance in the system due to pumping and losses via the atom and the cavity in the stationary state.
Here we used the atomic ground and excited state populations $1-P = (1-z)/2$ and $P = (1+z)/2$ and the photon number $N = |\alpha|^2$.
This is a universal relation independent of the particle's position and will be recovered several times throughout the paper.
Linking $P$ to $N$, Eq.\ \eqref{eq:cont} immediately yields the atomic population from the intracavity intensity which will appreciably simplify the analysis in the following.

Obviously for an atom fixed at a node of the cavity field where the atom-field coupling strength vanishes, the photon number is zero as well and the atomic upper state population is 
$P = \nu/(\nu+\gamma)$.
When the atom moves into regions where $G$ exceeds the threshold value $G_{th} = \sqrt{\kappa((\gamma+\nu)^2+\Delta^2)/(\nu-\gamma)}$,
the atom-field coupling opens an additional decay channel via the cavity mode.
Indeed, 
one can calculate the rate of emission into the cavity mode
\begin{equation}
  \label{eq:W_lamb}
   W = \frac{(\gamma+\nu)G^2}{(\gamma+\nu)^2 + \Delta^2} \, .
\end{equation}

\begin{figure}[t]
  \includegraphics[width=8cm]{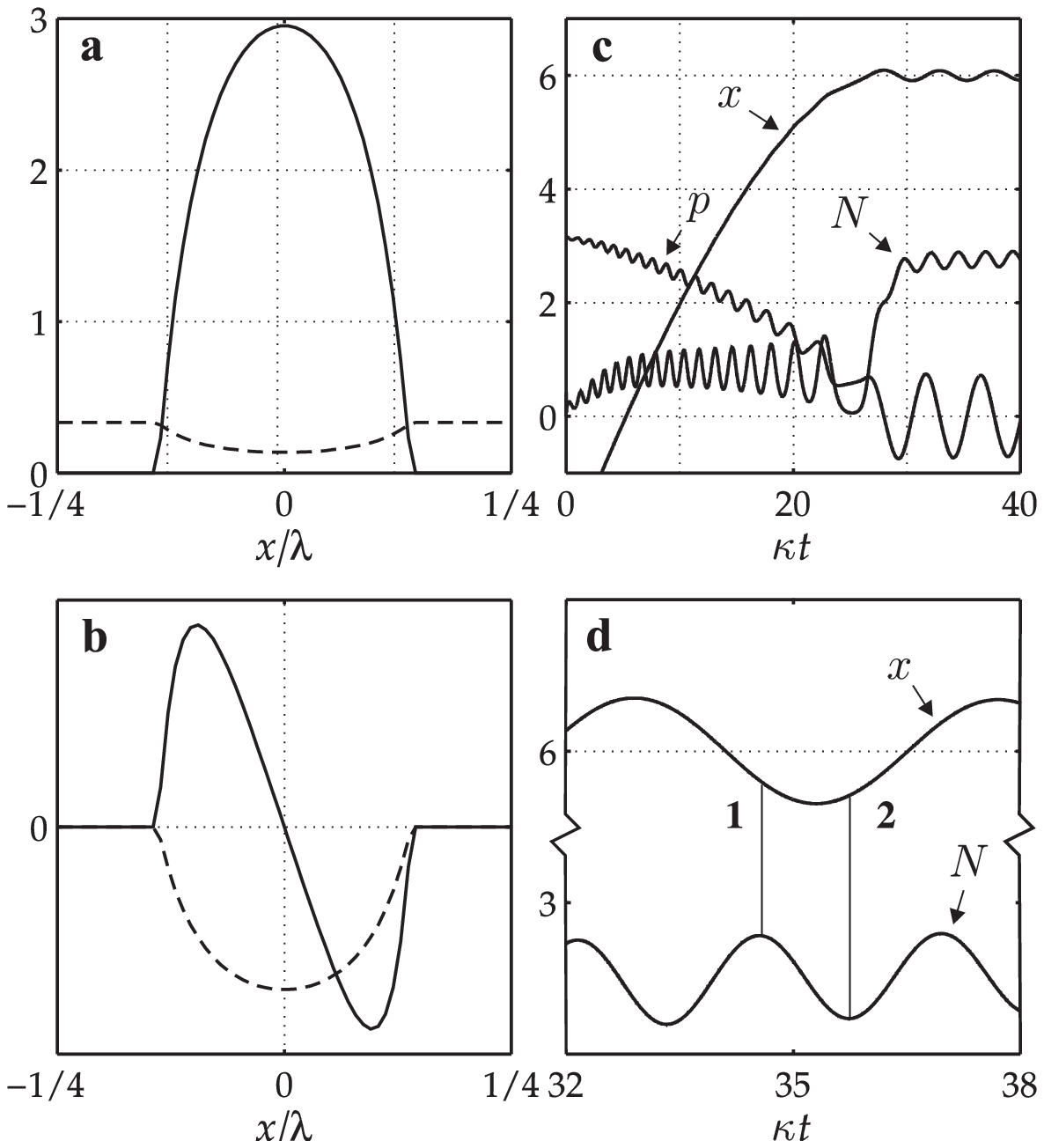}
  \caption{ \textbf{a} Photon number $N$ (solid line) and population inversion $z$ (dashed line) for the steady state as a function of the particle's position $x$. The parameters are $(\gamma,\nu,g,\Delta) = (10,20,100,200)\kappa$.\\
\textbf{b} Stationary force acting on the atom $F$ (solid line, a.\ u.) and corresponding potential $U$ (dashed line, a.\ u.) for the same parameters.\\ 
\textbf{c} Time evolution of the photon number $N$, particle position $x$ (units of $\lambda=2\pi/k$), and particle momentum $p$ (normalized to be unity when the atomic temperature equals the Doppler temperature $T= \hbar \gamma$ ) for the same parameters.\\ 
\textbf{d} Cutout of \textbf{c} demonstrating the cooling mechanism (see text). }
\label{fig:lamb}
\end{figure}

The general behavior of the cavity photon number is depicted in Fig.\ \ref{fig:lamb} \textbf{a} where we have plotted $N$ (solid line) as a function of the atomic position along the cavity axis within half a wavelength.
Note that the cavity field starts to be populated with $G$ crossing a threshold value in a highly nonlinear way.
According to Eq.\ \eqref{eq:cont}, this sudden increase has to be accompanied by a corresponding drop in the atomic population inversion $z = \kappa/W$ (dashed line).
At the same time there will be a big change of the light force on the atom, originating from the modified optical potential.
If the atom is a high field seeker, this already lets one expect a possible tight confinement of the atom.

For an atom at rest at a fixed position the mean force in steady state is simply proportional to the photon number as well as the gradient of the mode function and explicitly reads:
\begin{equation}
  \label{eq:F_lamb}
  F = \frac{2\kappa\Delta}{\gamma+\nu}\frac{\nabla G}{G}N\,.
\end{equation}
Obviously, $F$ will be zero at antinodes, where $N$ is maximal, due to the vanishing gradient of the mode function.
Notice that for $G\rightarrow 0$, $N$ tends faster to $0$ such that expression \eqref{eq:F_lamb} remains welll defined zero.
Since $W$, $z$, and $N$ are even functions of the detuning $\Delta$ above threshold, we get $F(-\Delta) = -F(\Delta)$, and the atom will be a high-field seeker for $\Delta > 0$.

This can be seen in Fig.\ \ref{fig:lamb} \textbf{b} which shows $F$ (solid line) as a function of $x$ for $\Delta = 200\kappa$ as well as the corresponding light potential, both in arbitrary units.
Although this sounds contradictory first when compared to standard formulas for the optical potential, one has to remember that the atom is inverted and thus the sign of the light potential is reversed and dominated by the upper level Stark shift.
Note that within this approximation a fixed atom will not feel any mean force unless $G > G_{th}$. 

Let us now look at the full coupled dynamics of atom and field by simultaneously integrating Eqs.\ \eqref{eq:ext} and \eqref{eq:lamb}, which can be easily performed numerically.
In Fig.\ \ref{fig:lamb} we show the evolution of the particle's position $x$ and momentum $p$ as well as the intracavity photon number $N$ for a typical set of parameters where one gets trapping.
The atom starts at some random position (plotted in units of $\lambda$) initially moving fast along the cavity axis.
Gradually, its motion gets damped until its kinetic energy falls below the potential depth, and the atom is then confined to oscillate in a single well.
Here its kinetic energy is still reduced further but at a much slower rate.
As a remarkable feature the photon number suddenly undergoes a drastic increase as the atom gets trapped.
This is due to the fact that the atom remains close to antinodes and never enters a spatial region, where the laser threshold is not fulfilled.
This could experimentally clearly be used to observe trapping in real time.

The physical mechanism responsible for the fast dissipation of the particle's motional energy during the initial stage is yet another variant of Sisyphus cooling.
Whenever the atom enters a spatial region where the system falls below threshold, the momentary photon number is higher than at the time later when it reenters the lasing region.
Hence, it is pulled back stronger during leaving than sucked in during reentering, which gives rise to net friction forces.
This type of friction force even continues while the particle is oscillating in a single well as illustrated in Fig.\ \ref{fig:lamb} \textbf{d}.
It shows a cutout of the trapping phase.
Due to the finite response time of the cavity field to the atomic position the cavity field attains a maximum value every time shortly after the atom has passed a field antinode (see \textbf{1}) and the atom still feels a significant friction.
On the other hand, when the atom approaches a turning point (\textbf{2}), the intensity reaches a minimum resulting in a smaller accelerating force towards the center.
This is of course very similar to passive cavity cooling in principle.
However, here the response time of the field is not only dominated by the cavity decay rate but by the full laser field dynamics.
This can strongly enhance the effective friction and lower cooling time and the steady-state temperature. 

\section{quantum rate equations for the internal dynamics}

Let us now go beyond the simple factorized c-number model and include fluctuations due to the interaction of the system with the environment represented by the noise operators in Eqs.\ \eqref{eq:hle_0}.
As one central consequence, these operators induce momentum diffusion of the atomic motion counteracting the cooling process and prevent the atom from stopping completely at a field antinode.
They also introduce fluctuations in the photon number and atomic occupation probabilities.

In the following quantum model we include the corresponding noise terms but we still assume a rather localized atomic wavepacket.
This allows to replace atomic momentum and position operators by their average values, i.e.\ we treat the atom like a classical Brownian particle in the optical potential.
However, we will keep the quantum correlations between the cavity field and the atomic polarization which were neglected in the previous section due to factorization.
We therefore base our treatment on second order operator products which turn out to obey a closed set of equations.
Using the abbreviations $\Phi = a^\dagger a$, $\Pi = \sigma_+\sigma_-$, $\Sigma = a^\dagger\sigma_- + \sigma_+a$, and
$\Lambda = (a^\dagger\sigma_- - \sigma_+a)/i$ we get:

\begin{subequations}
\label{eq:second}
  \begin{align}
    \dot{\Phi} = & -2\kappa\Phi + G\Sigma +\X_\Phi \label{eq:hle_n} \\
    \dot{\Pi} = & - 2\gamma\Pi - G\Sigma + 2\nu(1-\Pi)  + \X_\Pi \label{eq:hle_p} \\
    \dot{\Sigma} = & -\Gamma\Sigma - \Delta\Lambda + iG[\Sigma,\Lambda] + \X_\Sigma \label{eq:hle_s}\\
    \dot{\Lambda} = & -\Gamma\Lambda + \Delta\Sigma + \X_\Lambda \, .
  \end{align}
\end{subequations}
Here $\Gamma = \kappa + \gamma + \nu$ is the total damping rate of $\Sigma$ and $\Lambda$ that gives the atom field interaction energy and is closely related to the force.
The quantum fluctuations are contained in the operators $\X_i$ which again are fully determined by their second-order correlation functions given in appendix \ref{app:corr}.
Note that Eqs.\ \eqref{eq:second} are exact but still clearly nonlinear.
Thus, the corresponding equations for their expectation values are not closed and we have no explicit solution for their steady state.
The difficulties arise from the operator product in Eq.\ \eqref{eq:hle_s}.
An extra equation for $\langle\left[\Sigma,\Lambda\right]\rangle$ of course will inevitably incorporate higher order operator products
resulting in an infinite hirarchy of equations.
Following an idea developed in earlier laser models \cite{Protsenko}, we break this loop by replacing
\begin{equation}
  \label{eq:comm}
  i[\Sigma,\Lambda]  = 2(2\Pi-1)\Phi + 2\Pi \approx 2Z\Phi + 2\Pi\,,
\end{equation}
where $Z = \langle2\Pi-1\rangle$ is a real parameter that later can be calculated from Eq.\ \eqref{eq:cont} self-consistently.
This approximation means that we drop part of the quantum correlations between the atomic populations and the field intensity.
Fortunately, this turns out to play a minor role in the calculation of the system-variable expectation values in the parameter regime we are interested in.
The factorized equations for ${\bf \Xi} = (\Phi,\Pi,\Sigma,\Lambda)$ now read
\begin{equation}
\label{eq:lin}
  \frac{d}{dt}{\bf \Xi}  = \textbf{M}\,{\bf \Xi}  + \textbf{v} + {\bf \X}_\Xi \, ,
\end{equation}
where we have defined $\textbf{v} = (0,2\nu,0,0)$ and 
\begin{equation}
  \textbf{M} = \left(
    \begin{array}{cccc}
      -2\kappa & 0 & G & 0 \\
      0 & -2(\gamma+\nu) & -G & 0 \\
      2ZG& 2G & -\Gamma & -\Delta \\
      0 & 0 & \Delta & -\Gamma 
    \end{array}
  \right)\,.
\end{equation}
If the coupling strength $g$ is much less than the damping rate $\Gamma$ or the detuning $\Delta$, the operators $\Sigma$ and $\Lambda$ will adiabatically follow the values of $\Phi$ and $\Pi$ such that one is allowed to eliminate them.
Adiabatic elimination of $\Sigma$ and $\Lambda$ then yields the quantum rate equations for the photon number $N = \langle \Phi \rangle$ and the particle's upper state population $P = \langle \Pi \rangle$,
\begin{subequations}
  \label{eq:qre}
  \begin{align}
    \dot{N} &= -2(\kappa-ZW)N + 2WP \label{eq:qre_N} \\
   \dot{P} &= -2(\gamma+W+\nu)P - 2ZWN + 2\nu \, .\label{eq:qre_P}
\end{align}
\end{subequations}
Again,
\begin{equation}
  \label{eq:W_qre}
  W = \frac{\Gamma G^2}{\Gamma^2+\Delta^2}
\end{equation}
denotes the emission rate into the resonator.
In contrast to Eq.\ \eqref{eq:W_lamb}, $W$ now comprises the combined rate $\Gamma$.
It tends towards the rate found in the previous section when the cavity relaxation time $\kappa^{-1}$ is much longer than any other timescale of the system.

\begin{figure}
  \includegraphics[width=7.2cm]{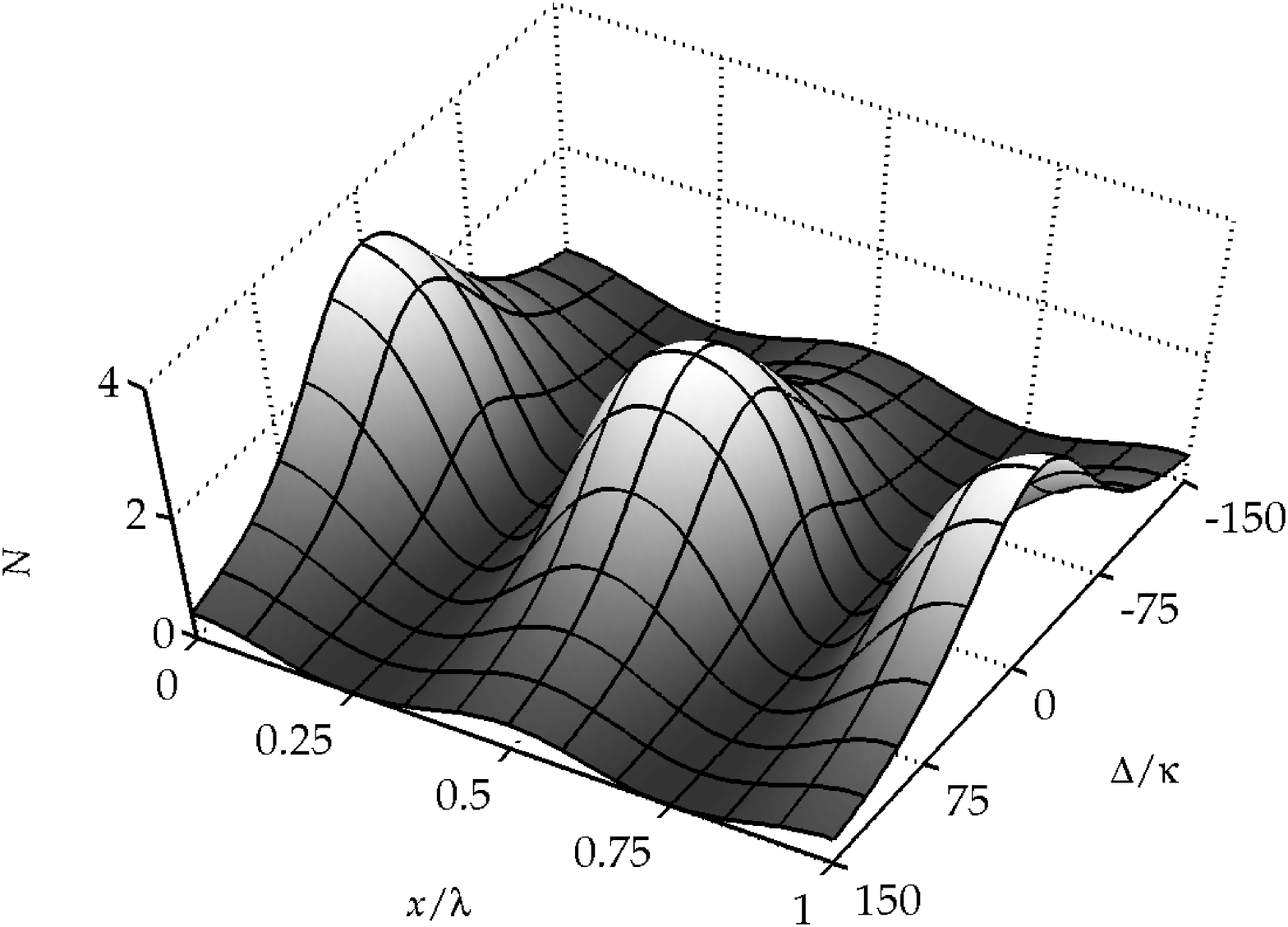}
  \includegraphics[width=7.2cm]{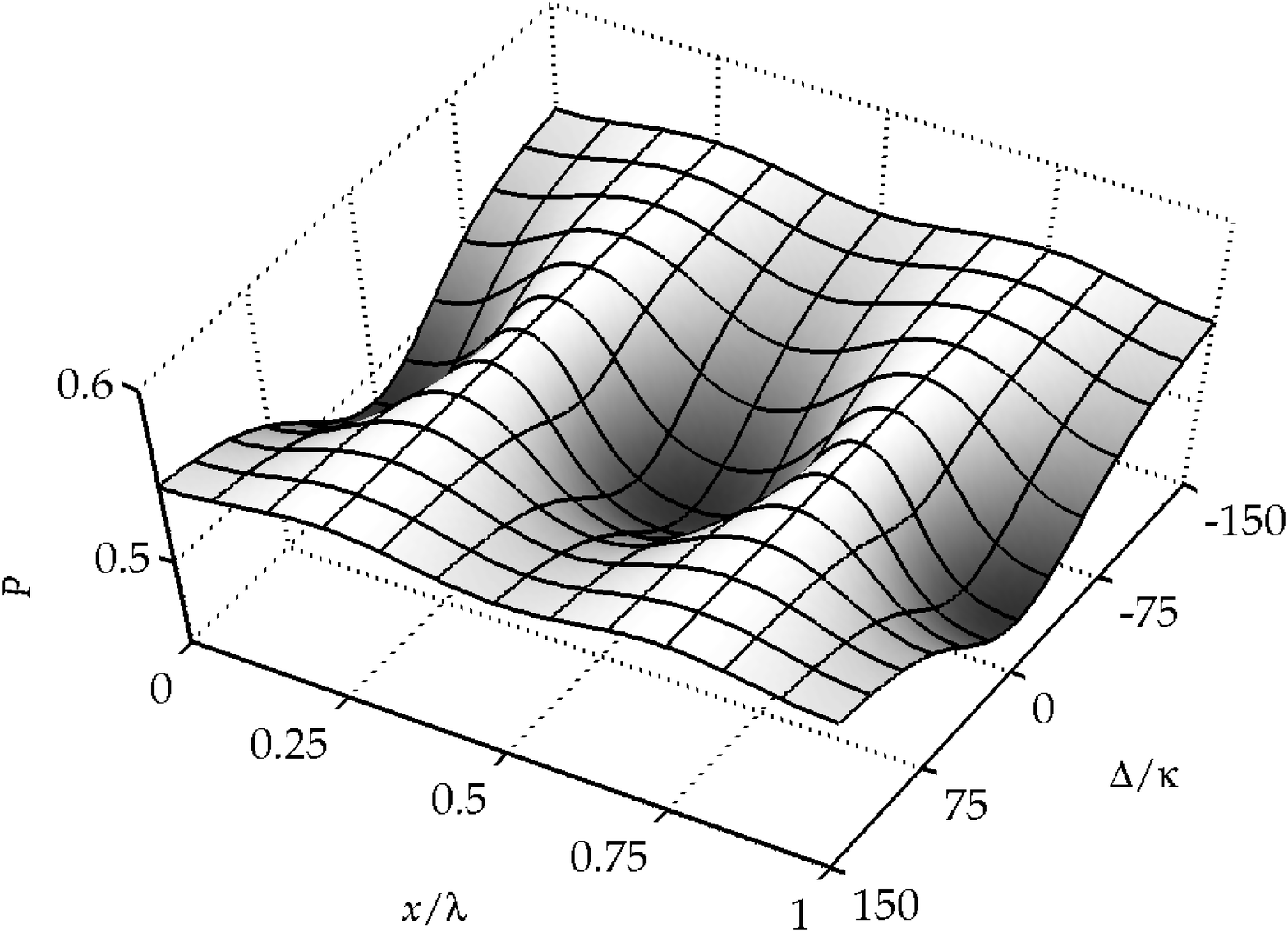}
  \includegraphics[width=7.2cm]{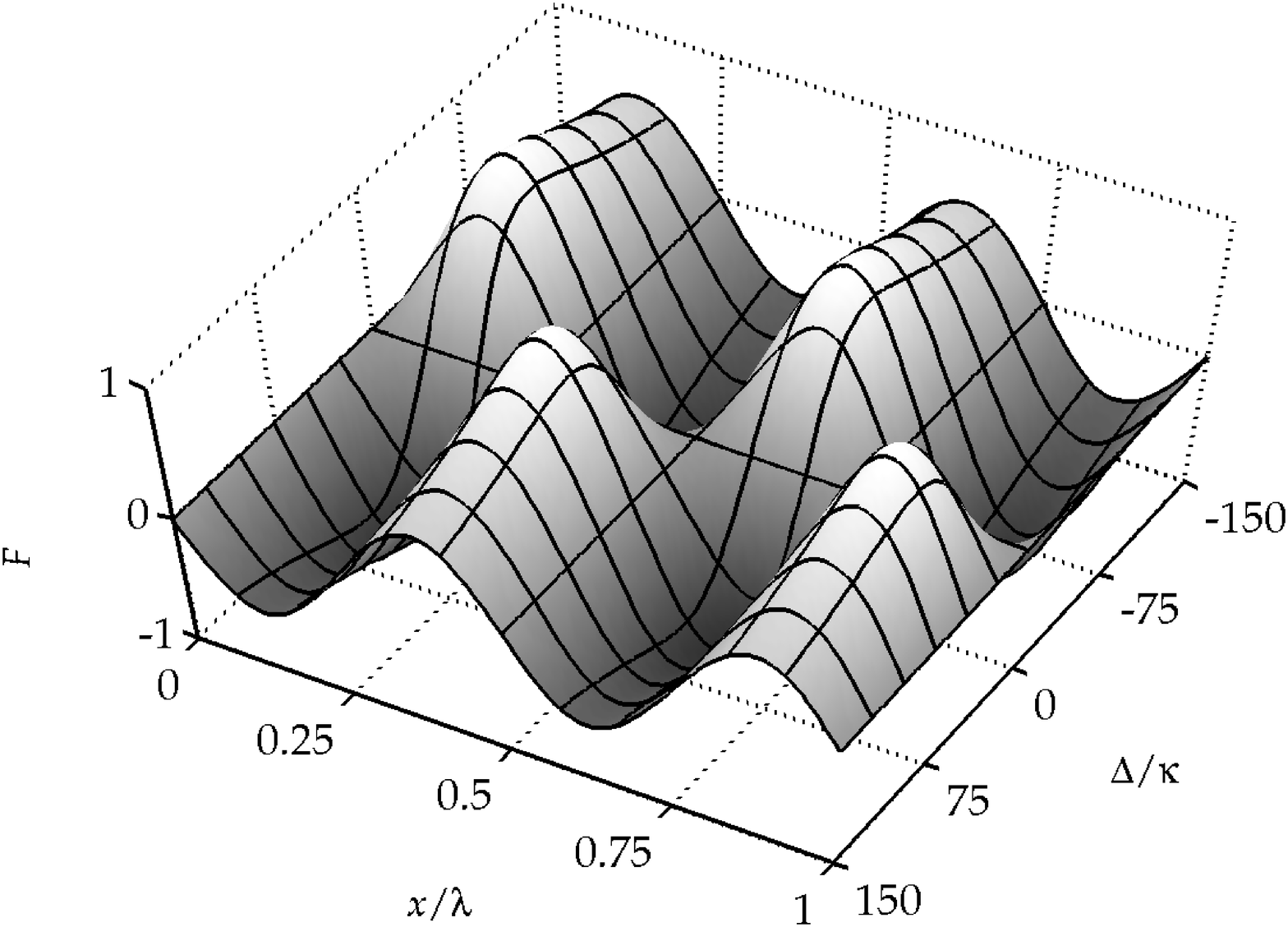}
  \caption{Up: stationary photon number obtained from the rate equations as a function of the atomic position
    $x$ and the detuning $\Delta$. The parameters are $(\gamma,\nu,g) = (20,25,20)\kappa$.\\
    Center: atomic upper state population for the same parameters.\\
    Down: The dipole force $F$ (plotted in arbitrary units) is an odd function of $\Delta$.}
\label{fig:sol_qre}
\end{figure}

Apart from relation \eqref{eq:cont} that immediately follows from the rate equations in steady state, there are two main effects arising from the strong atom-field coupling.
(Note that \eqref{eq:cont} is recovered whether or not the factorization \eqref{eq:comm} is made.)
First, the emission of photons into the resonator field is represented by the terms explicitely interlinking Eqs.\ \eqref{eq:qre}.
While the atomic population $P$ decreases, the field intensity grows due to the source term $2WP$.

Interestingly, both the atomic and the cavity linewidth are effectively modified as can be seen from the remaining terms proportional to $W$.
This intricately affects the decay properties of the whole system and, e.g.\ , reduces the resonator linewidth for an inverted atom.
As a result, even the stationary solution depends on the atomic position in a highly nonlinear way. 

Let us emphasize here that the cooling limit for conventional cavity cooling is usually related to the cavity linewidth $\kappa$ \cite{Hechenblaikner}.
An effectively gain reduced cavity damping rate thus can give rise to even lower final temperatures \cite{Vuletic}.
On the other hand, for an inverted atom momentum diffusion due to spontaneous emission is strongly pronounced and since $W$ can be of the order of $\gamma$, heating through dipole fluctuations will tend to raise the particle kinetic energy.
This effect can be expected to be reduced for several gain atoms in the mode where the inversion can be shared among many atoms. 

\section{Forces}

In the following we will investigate the atomic motion in more detail.
While the atom moves under the influence of the light forces induced by the cavity field, the atom modifies the light field dynamics according to its position.
For a passive resonator this mutual influence is a well-known feature of cavity QED and has been already seen experimentally \cite{Pinkse,Hood}.
There are further complications in the present system.
First, the atom itself generates the light field it interacts with.
Second, the photon creation from the incoherent pump is a highly nonlinear process exhibiting threshold.
In particular there is no light in the mode without an atom close to an antinode.
As the atom also provides gain inside the resonator, the lasing-induced trapping and cooling effects will be strongly enhanced when the system operates above threshold.

\subsection{Photons and forces in the steady state}

As is well known, the radiation pressure force on an atom at rest in a standing wave field cancels on average and only the dipole force is left \cite{Cohen}.
This remains true for a standing wave cavity field, and the only contribution to the net average force arises from the reactive response of the atom to the field dynamics (dipole force $F$).
For a slow enough atom the state of the field dynamically adjusts to its current position $x$ and can be well approximated by the steady state for an atom fixed at $x$.
In this adiabatic approximation we can calculate $F$ from the stationary expectation value of the force operator $\F$ that is given by the Heisenberg equation for the momentum operator $\cal \hat P$,
\begin{equation}
  \F = \frac{1}{i\hbar}[{\cal \hat P},{\H}] = -\nabla {\H} = (\nabla G)\Lambda\,.
\end{equation}
Inverting the matrix $\textbf{M}$ and defining $D = \det{\textbf{M}}/4$, we get from Eq.\ \eqref{eq:lin}:
\begin{equation}
  F = \frac{2\kappa\nu\Delta G}{D}(\nabla G) \, .
  \label{eq:F_qre}
\end{equation}
Here the environment is considered at zero temperature and hence the noise operators have vanishing expectation values and do not contribute.
From Eq.\ \eqref{eq:F_qre} we can read off that $F$ is an odd function of the detuning $\Delta$.
By help of the solution 
\begin{equation}
  N = \frac{\nu\Gamma G^2}{D}
  \label{eq:N_qre}
\end{equation}
for the photon number, we then   see that Eq.\ \eqref{eq:F_qre} tends to Eq.\ \eqref{eq:F_lamb} for $\kappa \ll \gamma + \nu$.
This corresponds to resonator fields close to coherent states with large mean intensity allowing a classical description.

In Fig.\ \ref{fig:sol_qre} we have plotted the stationary values of $N$, $P$, and $F$, respectively.
As expected from Eq.\ \eqref{eq:W_qre}, the photon number is a monotonic function of the atom-field coupling $G$ and decreases for growing mismatch between the atomic and resonator frequencies.
In contrast to the Lamb model, even a slight displacement of the atom from a field node causes the atom to radiate into the lasing mode although the system has not crossed the laser threshold.
This small but nonzero mode occupation leads to a force acting on the particle now.
Depending on the detuning, the atom is either pushed back to the node ($\Delta<0$) or attracted to the interaction region ($\Delta>0$).
The resulting light potential shows minima at antinodes, where we would like the atom to be trapped for positive detunings, and hence we will concentrate on this situation in the following.

\begin{figure}
  \includegraphics[width=8.6cm]{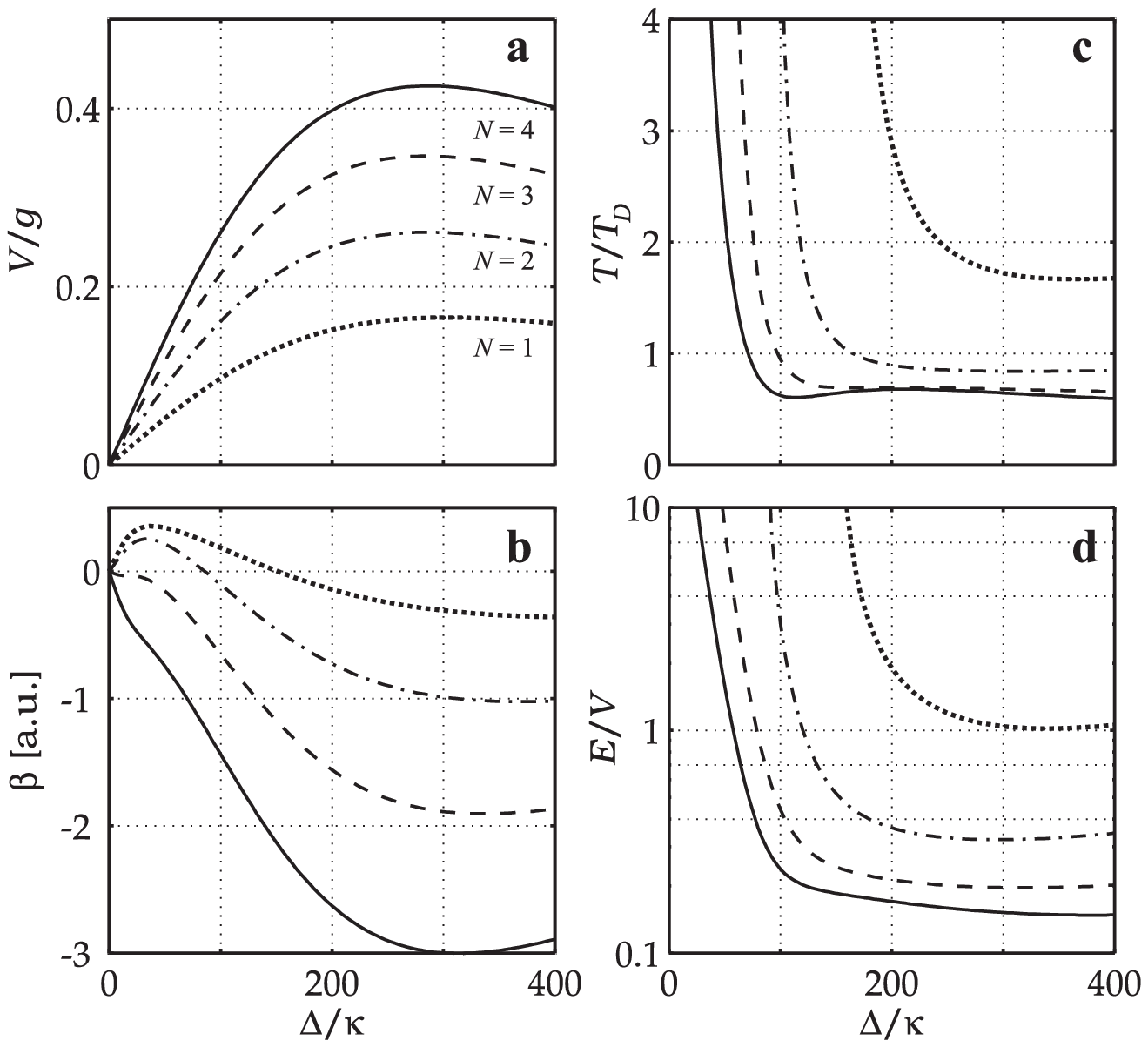}
  \caption{\textbf{a} Optical potential depth in units of the interaction energy for $(\gamma,g) = (10,50)\kappa$.
    In each curve the pump rate $\nu$ is adjusted resulting in constant maximum photon number $N$ as indicated.\\
    \textbf{b} Position averaged friction coefficient in arbitrary units for the same parameters as in \textbf{a}.\\
    \textbf{c} The atomic equilibrium temperature can be lower than the Doppler limit $T_D=\hbar\gamma$.\\
    \textbf{d} Ratio of atomic kinetic and potential energy.
    For $N>1$ it drops well below one when the cavity is far detuned from the atom indicating strong particle localization.}
  \label{fig:beta}
\end{figure}

Fig.\ \ref{fig:beta} \textbf{a} shows the potential depth $V$ in units of the interaction energy as a function of the detuning $\Delta$.
In each curve the pumping rate $\nu$ is adjusted in order to achieve constant maximum photon number $N$ as indicated.
The parameters are $(\gamma,g) = (10,50)\kappa$.
It is clear that $V$ successively increases with growing photon number.
On the other hand, large light intensities in the resonator will result in enhanced cooling and particle localization as we will see later.

\subsection{Friction and diffusion}

For finite particle velocities the field cannot follow the atomic motion instantaneously but will show some time-delayed response.
Let us now derive the linear velocity dependence of the force, i.e.\ the friction coefficient $\beta$.
If the atom moves much less than a wavelength before the internal variables attain their stationary values according to a displacement of the atom, one can derive the linearized correction term $\beta$ to the force as follows \cite{Gordon}.
Expanding the system operators in terms of the particle velocity $v$ and replacing the total time derivative by $\partial/\partial t + v\nabla$ we get a set of dynamical equations that can be solved systematically in different orders of $v$.
Writing $\langle{\bf\Xi}\rangle \approx {\bf X}^0 + v {\bf X}^1$, the zeroth- and first-order variables obey
\begin{subequations}
  \label{eq:orders}
  \begin{align}
    \mathbf{X}^0 &= -\textbf{M}^{-1} \textbf{v} \\
    \mathbf{X}^1 &= 
    -\textbf{M}^{-1} \nabla\left(\textbf{M}^{-1}\textbf{v}\right) \,, \label{eq:first}
  \end{align}
\end{subequations}
where the first line is precisely the afore discussed adiabatic solution.
The friction coefficient is then given by $\beta = (\nabla G)\langle \Lambda^1\rangle$ which is a rather lengthy expression and can be found in appendix \ref{app:fric}.
Note that in principle it is straight forward to derive the solution to higher orders in $v$, which would certainly produce more accurate results.
However, here we are mainly interested in the parameter regime with a very low final temperature of the atom.
In this limit the friction can be well approximated by the position-averaged force term linear in $v$. 

In order to give an estimate for the temperature, we also need the momentum diffusion coefficient $\cal D$ so that we can apply the Einstein relation \cite{Cohen}
\begin{equation}
  k_BT = {\overline \beta}/\overline{\cal D}\,.
  \label{eq:einstein}
\end{equation}
Hence, we also have to calculate the force fluctuations due to the coupling of the system to the vacuum modes.
In our case the cavity field as well as the atomic variables experience fluctuations around their stationary values which directly relate to force fluctuations.
This counteracts the cooling process and prevents the atom from stopping completely at antinodes.
Eq.\ \eqref{eq:einstein} then describes a situation where the contributions of friction and heating cancel and the atom reaches an equilibrium momentum distribution.

Similar to Brownian motion we calculate the diffusion coefficient $\cal D$ from the linear growth term of the momentum spread due to field fluctuations \cite{Cohen}.
Here we use the approach first outlined in Ref.\ \cite{Domokos2} which allows to approximately read off $\cal D$ from the two-time covariance of the force operator,
\begin{equation}
  \langle \F(t)\F(t-\tau)\rangle-\langle \F(t)\rangle\langle \F(t-\tau)\rangle = 2{\cal D} \delta(\tau) \, .
  \label{eq:D}
\end{equation}
A more detailed derivation of $\cal D$ can be found in appendix \ref{app:diff}.
Here we only want to note that, in addition to fluctuations of the cavity field, we have to consider momentum diffusion owning to the random recoil of spontaneously emitted photons and therefore have to add ${\cal D}_{\mbox{\tiny rec}}$ that can be found elsewhere \cite{Cohen}.
Note that in principle force fluctuations from the pumping mechanism would enter here as well.
As we do not explicitly specify the corresponding mechanism and assume that the pumping occurs transversally, we will neglect this contribution at this point.

\begin{figure}[t]
  \includegraphics[width=5.5cm]{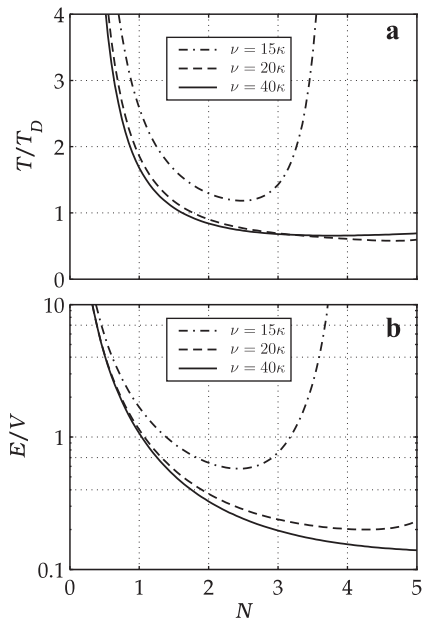}
  \caption{\textbf{a} Atomic temperature as a function of the maximum photon number $N$ in the cavity
    for different values of the pumping rate $\nu$.
    The other parameters are $(\gamma,g) = (10,50)\kappa$ while the atom-field detuning $\Delta$ was continuously changed.\\
    \textbf{b} Corresponding ratio $E/V$.}
  \label{fig:t_vgl}
\end{figure}

Let us now come to some numerical examples.
In Fig.\ \ref{fig:beta} \textbf{b} we have plotted the position-averaged friction coefficient for the same parameters as in \textbf{a}.
For very low intracavity fields we have $\beta \approx 0$ and the atomic motion is slightly accelerated rather than damped.
Only above threshold a strong friction force arises.
We see that for a given photon number we can have both heating and cooling, which already indicates that the system is active.
In the regime where $\nu<\gamma$ the atomic population is not yet inverted and $\beta > 0$.
For higher pumping strengths (i.e.\ larger $\Delta$ in Fig.\ \ref{fig:beta}) there appears a change in the signs of the population inverstion as well as of the friction coefficient.
It turns out that $Z>0$ is, together with $\Delta>0$, a main condition to achieve cooling and, moreover, large atomic upper state populations imply low temperatures as we will see at the end of this section.

Similar to the light potential  also $|\beta|$ shows a nonlinear increase with the photon number.
We can therefore expect the cooling efficiency to be strongly enhanced for higher laser intensities.
Indeed, the mean kinetic energy of the atom continuously decreases when more and more photons are present in the resonator mode.
This is demonstrated in Fig.\ \ref{fig:beta} \textbf{c} where we have plotted the atomic steady-state temperature in units of the Doppler temperature $T_D = \hbar\gamma$.
We clearly get sub-Doppler cooling in the cavity field, which definitely proves the important role of the cavity for cooling and is the prerequisite to combine trapping and cooling.  

So far we have seen that the atomic motion can be efficiently cooled when the atom is able to scatter ample photons into the resonator.
An essential issue to achieve long term operation of such a device is strong particle localization at antinodes.
Fig.\ \ref{fig:beta} \textbf{d} shows the ratio of the atomic kinetic energy $E$ and the optical potential $V$.
Below threshold, this ratio is much larger than unity and the particle's position is almost evenly distributed along the cavity axis.
For higher photon numbers, $E/V$ can drop well below unity corresponding to strong localization. This is in big contrast to free space Doppler cooling, where the average kinetic energy is shown to be always larger than the optical potential depth.

\begin{figure}[t]
  \includegraphics[width=6.5cm]{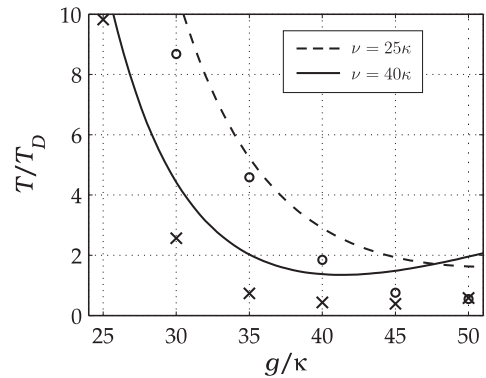}
  \caption{Atomic temperature as a function of the atom-field coupling constant $g$ for different values of the
    pumping rate where $(\gamma,\Delta) = (5,250)\kappa$.
    The marks show the results obtained from Monte Carlo wavefunction simulations.}
  \label{fig:t}
\end{figure}

Like in most other cavity cooling schemes, the pumping strength has great influence on the trapping and cooling rates.
Large light intensities induce fast and strong localization while the final temperature remains mostly unaffected there.
However, in our system where the atom acts like a gain medium inside the resonator, also the particle's kinetic energy shows strong dependence on the intracavity intensity.
This is demonstrated in Fig.\ \ref{fig:t_vgl} depicting the particle temperature and the ratio $E/V$ vs maximum photon number $N$ for different values of $\nu$.
The large correlation of the internal dynamics and the atomic motion leads to situations where the atom is glued to antinodes, thereby radiating light into the resonator mode, which in turn carries away energy and entropy from the system via the cavity mirrors and thus decreases the atomic temperature.

From Figs.\ \ref{fig:beta} and \ref{fig:t_vgl} we see that our equations predict lasing, cooling, and trapping simultaneously.
This occurs particularly for high photon numbers in the resonator and when the atom is far red detuned from the lasing mode.
In this parameter range, though, the atomic upper state has to be strongly populated and a large pumping rate $\nu$ is required, which appears to be the central experimental bottleneck in this system.
Naturally, this suggests to simultaneously use two or more atoms for gain.
In this way they can collectively emit into the lasing mode resulting in enlarged mode occupation without such stringent pumping requirements.
We thus expect not only the threshold appear at lower pumping strengths but also advanced cooling and trapping.

\begin{figure}[t]
  \includegraphics[width=8.6cm]{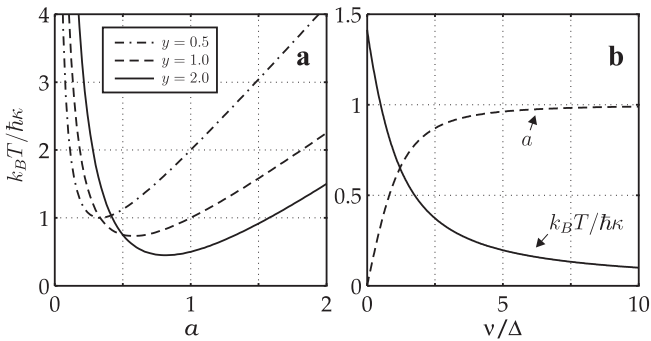}
  \caption{\textbf{a} Atomic temperature in units of $\hbar\kappa$.
    For $y>1/2$ the temperature can fall below one which corresponds to the limit of passive cavity cooling.\\
    \textbf{b} Minimum temperature as a function of $y = \nu/\Delta$ and the parameter $a$ determining the operating point of the system.}
  \label{fig:servus}
\end{figure}

\subsection{Comparison with numerical Monte Carlo simulations}

Since the early days of quantum optics, much attention has been paid on the development of laser theories at different levels of sophistication.
While most of the analytic models in the field of cavity-QED were based on rate equations at the beginning \cite{Yokoyama}, there has also been some work pointing out the shortcoming of the factorization approximation.
More accurate models were introduced \cite{Rice,Protsenko} that could be applied in our case as well.
We will, however, use a different approach here and directly use numerical methods to solve our original master equation without approximations.
Some first results on this were already published previously \cite{Salzburger04}.
Here we show a comparison with our analytical calculations in Fig.\ \ref{fig:t}.
Obviously, in the regime where lasing together with cooling coexists we find a surprisingly good agreement of the steady-state temperature and field expectation values. 

\subsection{Good-cavity limit}

In an active system the cavity field response time is no longer simply given by the cavity decay $\kappa$ rate but gets dynamically modified.
As $\kappa$ gives a lower limit on the kinetic temperature of the atom for passive cavity cooling, one could also expect changes here \cite{Vuletic}.
In the following we will study this in more detail and calculate the equilibrium temperature in the limit of very small $\kappa$ (particularly $\kappa \ll \nu$) by expanding friction and diffusion to first order in the decay rate $\kappa$.
In order to keep operating conditions comparable, we scale the rate of photon emission into the resonator also linear in $\kappa$, i.e.
\begin{equation}
  W = a\kappa\,.
\end{equation}
Here the parameter $a$ determines the operating point of the laser.
Note the threshold condition $\kappa = W$ found in the classical model; hence $a > 1$ corresponds to the laser working above threshold.
To keep the final expression simple we further assume $\gamma \approx 0$.
This leads to the following rather simple expression for the atomic equilibrium temperature
\begin{equation}
  k_B T = \hbar\kappa\,\frac{2a^2 + (a-1)^2y^2}{2ay}
\end{equation}
which depends on $a$ and the amount of pumping ratio $y = \nu/\Delta$.
In Fig.\ \ref{fig:servus} \textbf{a} we have plotted $k_B T/\hbar\kappa$ for different values of $y$.
We find that for $y > 1/2$ it can drop below one and thus below the limit of passive cavity cooling.
The respective minimum temperature,
\begin{equation}
  k_B T = \hbar\kappa\left(\sqrt{y^2+2}-y\right)\,,
\end{equation}
a monotonic decreasing function of $y$, is displayed in Fig.\ \ref{fig:servus} \textbf{b} (solid line).
Again large pumping rates and hence large values of $Z$ result in low temperatures.
In addition we show the corresponding $a$-parameter that remains slightly below one.
Therefore the system operates a bit below threshold where the atom is mainly in the excited state and the interaction energy is very large.

\section{conclusions}
We presented a simple self-consistent analytical model for the coupled dynamics of an inverted atom described as a point particle moving in the field of an single-mode resonator.
In good agreement with previously obtained predictions from Monte Carlo simulations we find that lasing, trapping, and cooling can simultaneously occur in such a setup, when the light mode is blue detuned from the atomic transition frequency and the pumping is sufficiently strong.
This surprising result turns out to be closely related to the fact that an inverted atom is a high field seeker for blue detuning.
Blue detuning is also a necessary condition for cooling as the missing photon energy in the stimulated emission process has to be taken from the atomic kinetic energy.
Luckily, this dissipation of kinetic energy via the resonator mode results in atomic equilibrium temperatures well below the Doppler limit and overcompensates the extra heating from the increased spontaneous emission of an inverted atom. 

As a consequence high photon numbers not only imply stronger localization of the atom but also lower temperatures.
As an extra bonus the atom as a gain medium effectively can reduce the resonator field linewidth below the cavity linewidth, so that under favorable conditions temperatures even below the limit of conventional cavity cooling ($k_B T = \hbar\kappa$) are possible.
This effect should definitely get more prominent for a larger atom number in the cavity.
Hence, even for larger samples stimulated cooling could be connected with lasing in a combined atom-laser--photon-laser setup providing for a coherent atomic beam and light source.

Let us finally remark that the fact that lasing is not necessarily connected to heating of the active medium but rather involves cooling could also prove important in rather different micro laser setups, e.g.\ on microchips, where heat production is a major issue preventing future miniaturization.

\acknowledgments
This work was supported by the Austrian Science Foundation FWF under projects P13435 and SFB ``Control and Measurement of Coherent Quantum Systems''.
P.\ D.\ acknowledges support from the National Scientific Fund of Hungary (Contract Nos.\ T043079, T049234), and the Bolyai Programme of
the Hungarian Academy of Sciences.

\begin{widetext}

\appendix

\section{Correlation functions of the noise operators $\Upsilon_i$}
\label{app:corr}
Analogous to the noise terms $\xi_i$ in Eqs.\ \eqref{eq:hle_0}, the operators $\Upsilon_i$ contain only free input field operators
and their expectation values vanish when evaluated at zero temperature.
The remaining non-zero correlation functions are
\begin{gather}
    \langle \X_\Phi(t) \X_\Phi(t-\tau) \rangle = 2\kappa N\,\del \nonumber \\
    \langle \X_\Pi(t) \X_\Pi(t-\tau) \rangle = \bigl(2\gamma P + 2\nu(1-P)\bigr)\,\del \nonumber \\ 
    \langle \X_\Sigma(t) \X_\Sigma(t-\tau) \rangle = \bigl(2\kappa P + 2\gamma N + 2\nu(1+N)\bigr)\,\del \nonumber \\
    \langle\X_\Lambda(t)\X_\Lambda(t-\tau)\rangle = \langle \X_\Sigma(t)\X_\Sigma(t-\tau)\rangle \nonumber \\
    \langle \X_\Phi(t) \X_\Pi(t-\tau)\rangle = \langle \X_\Pi(t)\X_\Phi(t-\tau)\rangle = 0 \nonumber \\
    \langle \X_\Phi(t)\X_\Sigma(t-\tau) + \X_\Sigma(t)\X_\Phi(t-\tau)\rangle = 2\kappa \langle\Sigma\rangle \,\del \nonumber \\
    \langle \X_\Phi(t)\X_\Lambda(t-\tau) + \X_\Lambda(t)\X_\Phi(t-\tau)\rangle = 2\kappa \langle\Lambda\rangle \,\del \nonumber \\
    \langle \X_\Pi(t)\X_\Sigma(t-\tau) + \X_\Sigma(t)\X_\Pi(t-\tau)\rangle = 2(\gamma-\nu) \langle\Sigma\rangle \,\del \nonumber \\
    \langle \X_\Pi(t)\X_\Lambda(t-\tau) + \X_\Lambda(t)\X_\Pi(t-\tau)\rangle = 2(\gamma-\nu) \langle\Lambda\rangle \,\del \nonumber \\
    \langle \X_\Sigma(t)\X_\Lambda(t-\tau) + \X_\Lambda(t)\X_\Sigma(t-\tau)\rangle = 0 \nonumber\,.
\end{gather}

\section{friction coefficient}
\label{app:fric}
From the solution of Eqs.\ \eqref{eq:orders} we obtain the somewhat unhandy expression
\begin{align}
  \beta = \frac{\nu\Delta(\nabla G)}{D^3}\biggl[&-G^3\Gamma\Bigl(4\kappa^2\left(\gamma+\nu\right)^2\Gamma +
    G^2\left(\gamma+\nu-\kappa\right)\left(\kappa^2+(\gamma+\nu)^2Z\right)\Bigr)\bigl(\nabla Z\bigr) \nonumber \\
    &+ 2\kappa\Bigl\{\left(\Gamma^2+\Delta^2\right)\left(G^2\left(\kappa^3-(\gamma+\nu)^3 Z\right)
    - 2\kappa^2\left(\gamma+\nu\right)^2\Gamma\right) \nonumber \\
    &\qquad + \Gamma G^4\bigl(\kappa-(\gamma+\nu)Z\bigr)^2
    + 2\kappa\left(\gamma+\nu\right)\Gamma^2 G^2 \bigl(\kappa-(\gamma+\nu)Z\bigr)\Bigr\}\bigl(\nabla G\bigr) \biggr] \, .
\end{align}

\section{diffusion coefficient}
\label{app:diff}
In the following we give a brief description of the calculation of the diffusion coefficient.
Writing $\F = \langle\F\rangle + \Upsilon$, definition \eqref{eq:D} yields
\begin{equation}
  2{\cal D}\,\del = (\nabla G)^2 \langle \Upsilon(t)\Upsilon(t-\tau)\rangle
\end{equation}
since $\langle\Upsilon(t)\rangle = 0$.
Here we see that the momentum spread directly arises from the noise exhibited by the interaction of the system with the environment
via the operator $\Upsilon$ that, for quasi-stationary conditions, is given by
\begin{equation}
  \Upsilon = \frac{1}{D}\Bigl((\gamma+\nu)\Delta G Z\,\Upsilon_\Phi + \kappa\Delta G\,\Upsilon_\Pi
    + \kappa(\gamma+\nu)\Delta \,\Upsilon_\Sigma 
    + \kappa(\gamma+\nu)\Gamma\,\Upsilon_\Lambda + G^2(\kappa-(\gamma+\nu)Z)\,\Upsilon_\Lambda\Bigr)\,.
\end{equation}
Inserting the correlation functions listed in appendix \ref{app:corr}, we find
  \begin{align} 
    {\cal D} = \frac{\nu G^2 (\nabla G)^2}{D^3}\biggl[
      &2\kappa^2\Delta^2\bigl(\gamma-\nu+(\gamma+\nu)Z\bigr)\Bigl(2\kappa(\gamma+\nu)\Gamma + G^2\bigl(\kappa-(\gamma+\nu)Z\bigr)\Bigr)\nonumber \\
      &+\Gamma^2\Bigl(\kappa^2(\gamma+\nu)^2\Delta^2 + \left[\kappa(\gamma+\nu)\Gamma + G^2\bigl(\kappa-(\gamma+\nu)Z\bigr)\right]^2\Bigr)\left(1+\frac{\kappa}{W}-Z\right)\nonumber \\
      &+\kappa\Gamma\Delta^2G^2\left(2\kappa\gamma\left(\frac{\kappa}{W}-Z\right)+\kappa^2 +(\gamma+\nu)^2Z^2\right)\biggr]\,.
  \end{align}

\end{widetext}

\end{document}